\documentclass[prc,aps,superscriptaddress,floatfix,twocolumn,nofootinbib]{revtex4-2}
\usepackage{graphicx}
\usepackage{makecell}
\usepackage{bm}
\usepackage{color}
\usepackage{ulem}
\usepackage{tikz}
\usepackage{ulem}
\newcommand\redsout{\bgroup\markoverwith{\textcolor{blue}{\rule[0.5ex]{2pt}{1.4pt}}}\ULon}

\renewcommand{\emph}[1]{\textit{#1}}

\begin{document}

\title{Alpha-Particle Monopole Form Factors with {\itshape Ab Initio} No-Core Shell Model}
\author{P.~Yin}
\thanks{Corresponding author: pengyin@iastate.edu}
\affiliation{College of Physics and Engineering, Henan University of Science and Technology, Luoyang 471023, China}
\affiliation{CAS Key Laboratory of High Precision Nuclear Spectroscopy, Institute of Modern Physics, Chinese Academy of
Sciences, Lanzhou 730000, China}
\affiliation{Department of Physics and Astronomy, Iowa State University, Ames, Iowa 50011, USA}

\author{A.~M.~Shirokov}
\affiliation{Skobeltsyn Institute of Nuclear Physics, Lomonosov Moscow State University, Moscow 119991, Russia}

\author{H.~Li}
\thanks{Corresponding author: lihe2007@impcas.ac.cn}
\affiliation{CAS Key Laboratory of High Precision Nuclear Spectroscopy, Institute of Modern Physics, Chinese Academy of
Sciences, Lanzhou 730000, China}
\affiliation{School of Nuclear Science and Technology, University of Chinese Academy of Sciences, Beijing 100049, China}

\author{B.~Zhou}
\affiliation{Key Laboratory of Nuclear Physics and Ion-Beam Application (MoE), Institute of Modern Physics, Fudan University,  Shanghai 200433, China}
\affiliation{Shanghai Research Center for Theoretical Nuclear Physics, NSFC and Fudan University, Shanghai 200438, China}

\author{X.~Zhao}
\affiliation{CAS Key Laboratory of High Precision Nuclear Spectroscopy, Institute of Modern Physics, Chinese Academy of
Sciences, Lanzhou 730000, China}
\affiliation{School of Nuclear Science and Technology, University of Chinese Academy of Sciences, Beijing 100049, China}

\author{S.~Bacca}
\affiliation{Institut f\"{u}r Kernphysik and PRISMA$^+$ Cluster of Excellence, Johannes Gutenberg-Universit\"{a}t Mainz, 55099 Mainz, Germany}

\author{J.~P.~Vary}
\affiliation{Department of Physics and Astronomy, Iowa State University, Ames, IA 50011, USA}

\begin{abstract}
The state-of-the-art {\itshape ab initio} nuclear many-body approaches with modern nuclear forces are challenged by the recent experimental measurement of the monopole form factor of the $0^+_1\rightarrow 0^+_2$ transition in the $\alpha$ particle [Kegel et al., Phys. Rev. Lett. \textbf{130}, 152502 (2023)]. We investigate
the elastic and inelastic $0^+_1\rightarrow 0^+_2$ transition form factors using
%this observable
%within the framework of
the \textit{ab initio} no-core shell model (NCSM). We observe a good convergence of both form factors with respect to the basis size employing the Daejeon16 nucleon-nucleon ($NN$) interaction. Our NCSM results are very close to the effective interaction hyperspherical harmonic calculations using $NN$ plus three-nucleon interactions based on the chiral effective field theory which take into account the continuum effects via the Lorentz integral transform. The significant difference between the {\itshape ab initio} results with various modern nuclear interactions and of some of them with
the recent experimental data provides motivations for deeper investigation of this observable.
\end{abstract}

 \maketitle

\textit{Introduction.}
The $\alpha$ particle, i.\ e., the $^4$He nucleus, is the lightest doubly magic nucleus and can be addressed as a natural laboratory for inspecting the inter-nucleon interactions and nuclear many-body methods~\cite{Kamada:2001tv}. The $^4$He ground state $0_1^+$ is deeply bound with the binding energy per nucleon of about $7$~MeV and has been extensively studied by nuclear physicists over the years. Its first excited state $0_2^+$ has the same spin and parity as the ground state. The $0_2^+$ state lies about $0.39$~MeV above the $p+^3$H threshold but about $0.37$~MeV below the opening of the $n+^3$He channel~\cite{Tilley:1992zz}. Therefore it is an unstable resonance decaying $100\%$ by proton emission.

Recently, the A1 Collaboration at Mainz Microtron (MAMI) has performed precision measurements of the monopole form factor of the $0^+_1\rightarrow 0^+_2$ transition via inelastic electron scattering~\cite{Kegel:2021jrh}. The new data improve the precision significantly compared to the previous measurements~\cite{Walcher:1970vkv,Frosch:1968sns,Kobschall:1983na}.

Several theoretical investigations have been performed for the transition form factor using realistic inter-nucleon interactions~\cite{Hiyama:2004nf,Bacca:2012xv,Michel:2023ley,Meissner:2023cvo,Viviani:2024yej}. In Ref.~\cite{Hiyama:2004nf}, a bound state technique, which expands the wave functions in a Gaussian basis, was used to calculate this observable with the simplified Argonne V8' (AV8') nucleon-nucleon ($NN$) interaction~\cite{Wiringa:1994wb} plus a phenomenological three-nucleon (3$N$) interaction. The results of Ref.~\cite{Hiyama:2004nf} describe relatively well %coincide well with
the experimental data. The effective interaction hyperspherical harmonic (EIHH) calculations in Ref.~\cite{Bacca:2012xv}, which take into account the continuum with the Lorentz integral transform (LIT), using the more accurate Argonne V$_{18}$ (AV18) $NN$ potential~\cite{Wiringa:1994wb} plus the Urbana IX (UIX) 3$N$ force~\cite{Pudliner:1995wk} and chiral effective field theory (EFT) based interactions ($NN$ interaction at N$^3$LO~\cite{Entem:2003ft} and 3$N$ interaction at N$^2$LO~\cite{Navratil:2007zn}) revealed sizable deviations from the experimental data. Accordingly, Ref.~\cite{Kegel:2021jrh} proposed that this scenario constitutes a puzzle: why do modern inter-nucleon interactions fail to explain the experimental data? As was pointed out in Ref.~\cite{Epelbaum:2023zz}, {\it ``the form factor may serve as a tool, or `magnifying glass', for probing small contributions to nuclear forces''}.

Following up the MAMI measurement, a calculation within
the no-core Gamow shell model in the coupled-channel representation (NCGSM-CC)~\cite{Michel:2023ley}, employing a $V_{low-k}$ version of the chiral $NN$ interaction at N$^3$LO~\cite{Entem:2003ft},  claimed to explain the experimental data by considering the continuum coupling. However, the study of Ref.~\cite{Michel:2023ley} neglected 3$N$ forces and revealed a strong sensitivity to  $\Delta E_{\rm thr}$, the position of the  $0^+_2$ resonance with respect to the $p-^3$H threshold. It was found that by fitting the transition form factor to the experimental inelastic form factor through varying $\Delta E_{\rm thr}$ instead of calculating $\Delta E_{\rm thr}$ within the NCGSM, a good agreement with the Kegel {\it et al.} data can be obtained at $\Delta E_{\rm thr}=0.12$ MeV~\cite[Erratum]{Michel:2023ley}  that is much smaller than the experimentally observed value $0.39$ MeV.

In Ref.~\cite{Meissner:2023cvo}, the {\itshape ab initio} nuclear lattice EFT  calculations, which is also a bound state technique, reproduced the experimental data, albeit with a simplified nuclear interaction formulated on a periodic cubic box~\cite{Lu:2018bat} which is not known to provide a high quality description of the $NN$ phase-shift data.
In addition, we note that the monopole transition density presented
in~\cite[Fig.~2]{Meissner:2023cvo} after the Fourier transform results in an about 20$\%$ higher form factor with respect to that presented in~\cite[Fig.~1]{Meissner:2023cvo}~\cite{Shen}, which raises a concern over the robustness of the claimed agreement with experiment.

The most recent results with the hyperspherical harmonic (HH) method in Ref.~\cite{Viviani:2024yej}, which includes the $p+$$^3$H and $n+$$^3$He scattering states, reproduced well the inelastic form factor experimental data by using the chiral interactions ($NN$ interaction at N$^3$LO~\cite{Entem:2003ft} together with 3$N$ interaction at N$^2$LO~\cite{Gardestig:2006hj,Gazit:2008ma}). However, this calculation results
in a form factor that significantly differs from that obtained in Ref.~\cite{Bacca:2012xv} using    the same $NN$ but another parametrization of the chiral 3$N$ force. The differences in the 3$N$ interaction parametrization are not expected to produce an approximately $50\%$ disagreement in the values of the transition form factor around its maximum.
This difference may result from the fact that the $p+^3$H channel allowed in Ref.~\cite{Viviani:2024yej} includes both isospins $T=0$ and $T=1$ thus suppressing the role of the $T=0$ component of the $0^+_2$ resonance, which provides a dominant contribution to the monopole transition, and therefore the transition form factor in the calculations of Ref.~\cite{Viviani:2024yej} is reduced. Interestingly, in Ref.~\cite{Viviani:2024yej} the position of the $0^+_2$ state is predicted to be about $0.1$~MeV above the $p+^3$H threshold,  much below the experimental value.

%The position of the $0^+_2$ state is predicted to be about $0.1$~MeV above the $p+^3$H threshold %in Ref.~\cite{Viviani:2024yej} (much below the experimental $0.39$~MeV), which implies that the %transition monopole form factor and the location of the $0_2^+$ may not be very correlated.

The issue of describing the inelastic monopole form factor using modern nuclear forces remains unresolved.
%Though some calculations claimed to reproduce the experimental data successfully, some important questions deserve more investigations. For example, the
The differences between various reported theoretical calculations and at least some
of them with
the MAMI experimental data %and some theoretical calculations
raise significant questions. %In the meantime, we noticed that various calculations adopting different interactions may lead to very different monopole transition form factor.
Additional studies of this observable could shed further light on the Hamiltonian and method dependences.

In this letter, we present an alternative {\itshape ab initio} calculation for the monopole form factor of the $0^+_1\rightarrow 0^+_2$ transition in $^4$He with the no-core shell model (NCSM).
We approximate the resonance as a bound state in the NCSM, which typically provides accurate results for narrow resonances~\cite{Li:2024kno,Caprio:2025osg,4He,Barrett:2013nh,Caprio:2019yxh,McCoy:2020xhp}, such as the case for the $0^+_2$ state of $^4$He (with about $0.5$ MeV width).
We perform the calculation with the Daejeon16 $NN$ interaction~\cite{Shirokov:2016ead}. This interaction is based on the N$^3$LO chiral EFT interaction of Ref.~\cite{Entem:2003ft}, softened via a similarity renormalization group (SRG) transformation (with the flow parameter $\lambda=1.5$ fm$^{-1}$) so as to improve convergence of \textit{ab initio} studies, modified off-shell to mimic three-nucleon forces and provide one of the best descriptions of nuclei with $A\le16$~\cite{Maris:2019etr}. Electromagnetic properties of nuclei have been successfully investigated with the Daejeon16 interaction~\cite{4He,Li:2024kno,Otsuka:2022bcf,Caprio:2022mkg,Caprio:2025osg}.

\textit{Theoretical methods.}
The monopole transition form factor $F(q^2)$ is related to the point-proton transition density $\rho_{\rm tr}(r)$ by
\begin{equation}
\left|F(q^2)\right|^2=\left|\frac{4\pi}{Z}\int\rho_{\rm tr}(r)j_0(qr)r^2 dr\right|^2 f_p^2(q^2),
\label{eq:FF}
\end{equation}
where $Z=2$ is the charge of $^4$He under consideration~\cite{Meissner:2023cvo}.
The proton form factor $f_p(q^2)$ takes care of the proton finite size. As was done in Ref.~\cite{Meissner:2023cvo}, we use a microscopic $f_p(q^2)$ from Ref.~\cite{Lin:2021xrc} corresponding to the proton charge radius of $0.84$ fm, which is very close to the most recent experimental data $0.84075(64)$ fm~\cite{Mohr:2024kco}. A phenomenological expression $f_p(q^2)=1/(1+0.0548q^2)^2$ ($q$ in fm$^{-1}$), which corresponds to the proton charge radius of $0.81$~fm, has been frequently used in calculations.
We found that the monopole form factors of $^4$He calculated with the microscopic and phenomenological proton form factors are nearly indistinguishable over the range of the experimental data.
The point-proton transition density $\rho_{\rm tr}(r)$ is the matrix element of the point-proton density operator $\hat{\rho}(\vec{r})=\sum_{i=1}^Z \delta(\vec{r}-\vec{r}_i)$, where $\vec{r}_i$ denotes the position vector of $i$th proton, between the ground state $0_1^+$ and the first excited $0_2^+$ state~\cite{Hiyama:2004nf,Meissner:2023cvo},
\begin{equation}
\rho_{\rm tr}(r)=\langle0_1^+|\hat{\rho}(\vec{r})|0_2^+\rangle.
\label{eq:CD}
\end{equation}

The transition form factor transforms to the elastic form factor if we replace the point-proton transition density $\rho_{\rm tr}$ in Eq.~(\ref{eq:FF}) by the ground state point-proton density. There are various definitions for the monopole transition form factor in available references. In order to compare with the most recent experimental data % in
of Ref.~\cite{Kegel:2021jrh} and make comparisons with other %among different
theoretical calculations, one should ensure that the corresponding elastic monopole form factors satisfy
\begin{equation}
|F_{\rm el}(q^2=0)|=1.
\label{eq:EFM}
\end{equation}

We use the \textit{ab initio} NCSM~\cite{Barrett:2013nh} to calculate the wave functions of the $0_1^+$ state and the $0_2^+$ state. The NCSM has been extensively used in studies of light nuclei %$s$- and $p$-shell nuclei
(see, e.\,g., Refs.~\cite{Maris:2016wrd,Maris:2020qne,LENPIC:2022cyu,7He,9Li,4He,3n,sd,Magic}).
In the NCSM, the nuclear wave functions are obtained by diagonalizing the chosen nuclear Hamiltonian in a truncated Slater determinant harmonic oscillator (HO) basis characterized by the basis oscillator parameter $\hbar\Omega$.  %In fact, we
The truncation of the model space is determined and labeled by the number of HO excitation quanta, $N_{\rm max}$, relative to the minimum number of quanta required by the Pauli principle. We test convergence by showing calculated quantities vs $N_{\rm max}$, and we report how these quantities approach their asymptotic values as $N_{\rm max}$ increases.

In this work, we perform NCSM calculations with the code MFDn using the Lanczos algorithm~\cite{Aktulga:2014im,Maris:2010im,Maris:2022im,Cook:2021im}. We use the Daejeon16 {\it NN} interaction~\cite{Shirokov:2016ead} and add the Coulomb interaction between the protons.
We adopt the Lawson method~\cite{Gloeckner:1974sst,Whitehead:1977qp} to ensure that only states which are free from spurious center of mass (c.m.) excitation
remain in the calculated spectrum. We employ the method in Ref.~\cite{Cockrell:2012vd} to obtain the translational invariant density in Eq.~(\ref{eq:FF}).

The $0_2^+$ state of $^4$He is a resonance sitting just above the $^3$H $+$ $p$ threshold.
Therefore the continuum state $0_2^+$ obtained with the NCSM in the above manner is viewed as based on a discretized approximation of the continuum. This approximation becomes more reliable as the basis size increases. For example, in Ref.~\cite{4He} we calculated the $E1$ polarizability, an inverse energy weighted sum rule, of $^4$He by directly solving for thousands of discretized $1^-$ states and reached robust convergence with $N_{\rm max}$ increasing.

\begin{figure}[t]%[hbtp]
%\begin{center}
%\includegraphics[width=0.6\textwidth]{Graph2.pdf}
\includegraphics[width=1.0\linewidth]{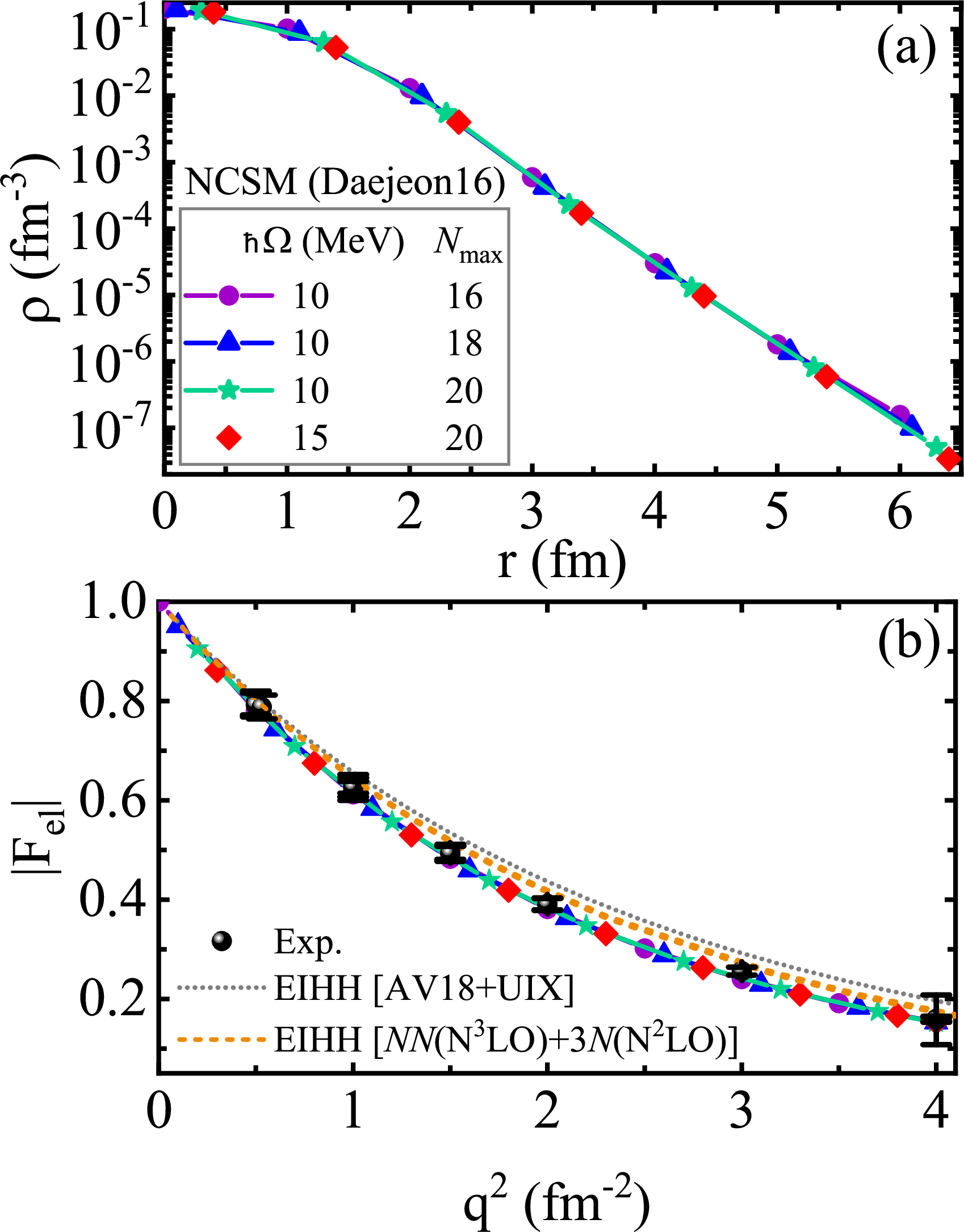}
%\end{center}
\caption{(Color online) Point-proton density [panel (a)] and elastic monopole form factor [panel (b)] of the $^4$He ground state calculated with the NCSM using the Daejeon16 $NN$ interaction for $\hbar\Omega=10$~MeV at $N_{\rm max}=16{-}20$ (line+symbol) and $\hbar\Omega=15$~MeV at $N_{\rm max}=20$ (red solid diamond). In panel (b), the experimental data along with the quoted uncertainty (black solid dots with error bars) are taken from Ref.~\cite{Frosch:1967pz}; EIHH results calculated with AV18+UIX and chiral $NN$ (N$^3$LO)+3$N$ (N$^2$LO) interactions from Ref.~\cite{Bacca:2012xv} are shown for comparison.}\label{fig1}
\end{figure}
\textit{Results and discussion.}
We start with calculations of the elastic $^{4}$He form factor.
In Fig.~\ref{fig1} (a) we present the point-proton density distribution of the $^4$He ground state $0_1^+$ for $N_{\rm max}=16{-}20$ with the same $\hbar\Omega=10$~MeV. We observe good exponential tail for the proton density and its robust convergence with respect to $N_{\rm max}$. The results for $N_{\rm max}=16{-}20$ are nearly indistinguishable in Fig.~\ref{fig1} (a). We performed similar calculations for $\hbar\Omega=15$~MeV and found also robust convergence. For simplicity we show in Fig.~\ref{fig1} (a) only the results of $\hbar\Omega=15$~MeV at $N_{\rm max}=20$. We find that the results of $\hbar\Omega=10$ and $15$~MeV at $N_{\rm max}=20$ are in good agreement with each other.

Based on the point-proton density distribution shown in Fig.~\ref{fig1} (a), we can obtain the elastic form factor straightforwardly as we mention above. In Fig.~\ref{fig1} (b) we present the elastic form factor of the $^4$He ground state $0_1^+$ for $N_{\rm max}=16{-}20$ with the same $\hbar\Omega=10$~MeV and for $N_{\rm max}=20$ with $\hbar\Omega=15$~MeV. As we explain in Eq.~(\ref{eq:EFM}), the elastic form factor is unity at $q^2=0$ fm$^{-2}$. The NCSM results show good convergence with respect to $N_{\rm max}$. The converged results are in good agreement with the experimental data, which is not surprising, since the Daejeon16 interaction gives the $^4$He point-proton radius of $1.514$~fm, which is very close to the experimental value of 1.484(5) fm. Our results are also in good agreement with the EIHH calculations with the Argonne V$_{18}$ (AV18) $NN$ potential plus the Urbana IX (UIX) 3$N$ force and Chiral interactions ($NN$ interaction at N$^3$LO and 3$N$ interaction at N$^2$LO) in Ref.~\cite{Bacca:2012xv}, which give the $^4$He point-proton radius of $1.432(2)$ and $1.464(2)$ fm, respectively. Neglecting two-body currents and further relativistic corrections, the experimental value of the point-proton radius $r_p$ is obtained using %with
the experimental charge radius $r_c=1.681\pm0.004$ fm~\cite{Sick:2008zza} via the following expression~\cite{Jentschura:2011}:
\begin{equation}
r_{p}^2=r_c^2-R_p^2-(N/Z)R_n^2-\frac{3}{4m_p^2} ,
\label{eq:rps}
\end{equation}
where $R_p=0.84075(64)$ fm~\cite{Mohr:2024kco} is the proton finite size, $R_n^2=-0.1161(22)$ fm$^2$~\cite{ParticleDataGroup:2022pth} is the neutron finite size correction, $m_p$ is the mass of the proton, and the last term is a relativistic correction.

We noticed that the point-proton density and the elastic form factor have been calculated with the NCSM using the Daejeon16 interaction at $N_{\rm max}=18$ in Ref.~\cite{Baker:2023wla}. The proton form factor adopted in Ref.~\cite{Baker:2023wla}, which corresponds to the proton charge radius of $0.863(4)$ fm, was taken from Ref.~\cite{Galynskii:2024thr}. Our results are consistent with those in Ref.~\cite{Baker:2023wla}. We illustrate the convergence of the these two quantities as the basis size increases.

The binding energy of $^4$He obtained with the Daejeon16 interaction is  $28.372(0)$ MeV~\cite{Shirokov:2016ead}. The energy of the $0_2^+$ state is $-8.377(56)$  MeV which is obtained by a simple $3$-point exponential extrapolation~\cite{Maris:2019etr,Maris:2008ax} using the calculated results at $N_{\rm max}=16, 18$ and $20$ and averaged over the best convergence range $\hbar\Omega=7-15$ MeV. The uncertainty is evaluated as half of the difference between the maximum and minimum extrapolated values over the above $\hbar\Omega$ range. The corresponding  excitation energy of the $0_2^+$ state $19.995(56)$ MeV is in good agreement with the experimental value $20.210$ MeV.

The robust convergence and good description of the experimental data demonstrated in Fig.~\ref{fig1} support the reliability of our calculations of the $0^+_1\rightarrow 0^+_2$ monopole transition form factor of $^4$He.

\begin{figure*}[htpb]
%\begin{center}
\includegraphics[width=\textwidth]{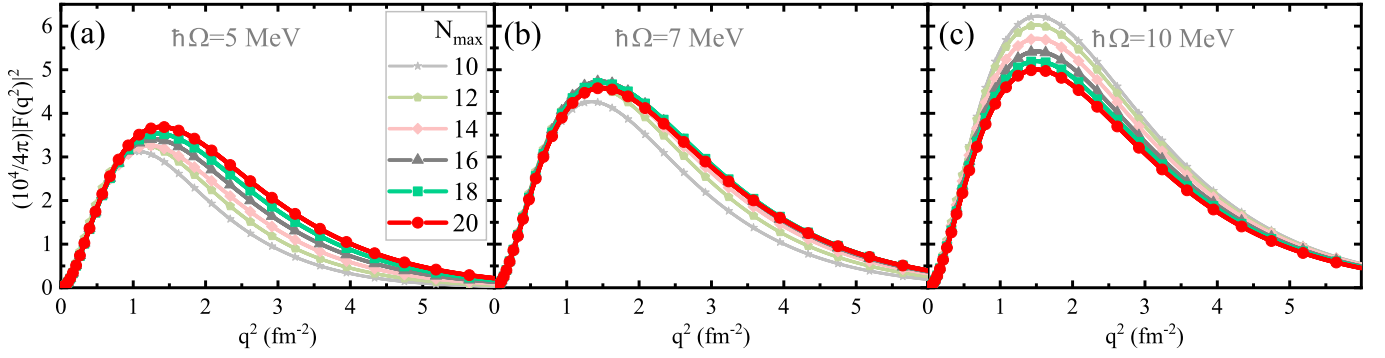}
%\end{center}
\caption{(Color online) Monopole form factor of the $0^+_1\rightarrow 0^+_2$
transition in $^4$He calculated with the NCSM using the Daejeon16 $NN$ interaction
for $\hbar\Omega=5$ [panel(a)], $7$ [panel (b)] and $10$ [panel (c)] MeV.}\label{fig3}
\end{figure*}

In Fig.~\ref{fig3} we show the calculated monopole form factor of the $0^+_1\rightarrow 0^+_2$
transition in $^4$He at $N_{\rm max}=10{-}20$ with $\hbar\Omega=5$, $7$ and $10$~MeV. We observe in Fig.~\ref{fig3} that the convergence of the transition monopole form factor is challenging for $\hbar\Omega=5$ and $10$~MeV. The results of $\hbar\Omega=7$~MeV show the best convergence. In particular, the results of $\hbar\Omega=7$~MeV at $N_{\rm max}=16, 18$ and $20$ are nearly indistinguishable. The results with $\hbar\Omega=5$ and $10$~MeV tend towards the converged results obtained with $\hbar\Omega=7$~MeV as $N_{\rm max}$ increases though both of these two cases do not reach apparent convergence up through $N_{\rm max}=20$ [see panels (a) and (c) in Fig.~\ref{fig3}].

\begin{figure}[htpb]
%\begin{center}
\includegraphics[width=\columnwidth]{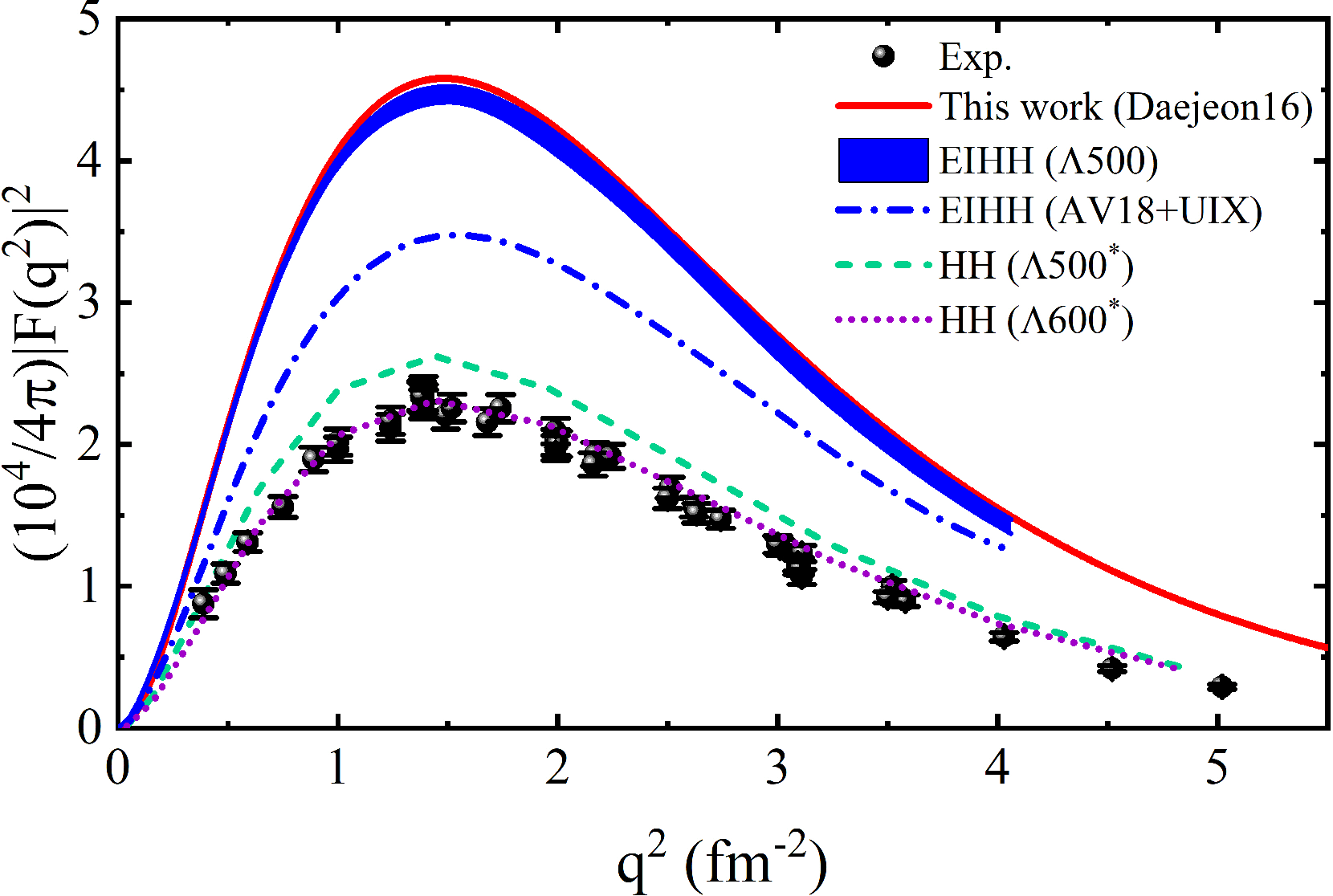}
%\end{center}
\caption{(Color online) Monopole form factor of the $0^+_1\rightarrow 0^+_2$
transition in $^4$He calculated with the NCSM using the Daejeon16 $NN$ interaction
for $\hbar\Omega=7$~MeV at $N_{\rm max}=20$ (solid line). The experimental data along with their quoted uncertainties (solid dots with error bars) from Ref.~\cite{Kegel:2021jrh}, the EIHH results
from Ref.~\cite{Bacca:2012xv}
obtained with the AV18+UIX interactions (dash-dotted line) and the chiral $NN$(N$^3$LO)\,+\,3$N$(N$^2$LO) interactions
adopting the cutoff $\Lambda=500$~MeV (blue band), and the HH results
from Ref.~\cite{Viviani:2024yej} obtained with chiral $NN$ (N$^3$LO) and differently parameterized 3$N$ (N$^2$LO)
%bare chiral $NN$ (N$^3$LO)+3$N$ (N$^2$LO)
%\remove{with the same}
interactions %from Ref.~\cite{Viviani:2024yej}
adopting %two
cutoffs $\Lambda=500$~MeV (dashed line, denoted as $\Lambda500^{*}$ in the legend) and $\Lambda=600$~MeV (dotted line, denoted as $\Lambda600^{*}$ in the legend) are shown for comparison.}\label{fig4}
\end{figure}
In Fig.~\ref{fig4} we compare our results with the MAMI experimental data. We plot the NCSM results of $\hbar\Omega=7$~MeV at $N_{\rm max}=20$. We also present in Fig.~\ref{fig4} the results of the state-of-the-art chiral interactions calculated with the HH method~\cite{Viviani:2024yej} and the EIHH method~\cite{Bacca:2012xv}. The AV18+UIX results from Ref.~\cite{Bacca:2012xv} are also shown in Fig.~\ref{fig4} for comparison.
Our result (red curve in Fig.~\ref{fig4}) is very close to the EIHH calculation with the chiral interactions (blue band in Fig.~\ref{fig4}), both of which are not able to explain the experimental data. In Ref.~\cite{Bacca:2012xv}, the EIHH calculation takes into account the continuum effects for the $0_2^+$ state via the LIT method.
We note that the Daejeon16 interaction used in our study originates from the same N$^3$LO interaction utilized in Ref.~\cite{Bacca:2012xv} while the role of the 3$N$ interaction is mimicked by the off-shell properties of the Daejeon16 interaction. The excellent agreement of our NCSM results with those of Ref.~\cite{Bacca:2012xv} suggests that the coupling to the continuum may not be essential for the transition monopole form factor of $^4$He.

We noticed that, in Ref.~\cite[Erratum]{Michel:2023ley}, the Daejeon16 interaction was also used to calculate the transition monopole form factor of $^4$He with the NCGSM-CC method, which shows quite different results from our NCSM calculation. However, one should note that the results of Ref.~\cite[Erratum]{Michel:2023ley} were obtained by tuning the parameter $\Delta E_{\rm thr}$ to fit the experimental data on the transition form factor.
%Therefore it may not be valuable to compare the results of Ref.~\cite{Michel:2023ley} with {\it ab initio}  calculations.
%
We do not show also in Fig.~\ref{fig4} the results from
Refs.~\cite{Hiyama:2004nf,Meissner:2023cvo} which reproduce experimental data with simplified interactions.
%One should note that the monopole transition form factor and the transition density plotted in Ref.~\cite{Meissner:2023cvo} are not self-consistent.

\vspace{1mm}
\textit{Conclusions and outlook.}
In conclusion, we investigated the monopole form factors in $^4$He with the {\itshape ab initio} NCSM. We performed calculations with the realistic Daejeon16 $NN$ interaction.
The translational invariant density and the elastic form factor of the ground state $0^+_1$ are found to be well converged as the basis size increases. The elastic form factor we calculated is in good agreement with the experimental data and the EIHH calculations with the AV18 + UIX potential and chiral $NN$+3$N$ interactions.
Our calculated monopole form factor of the $0^+_1\rightarrow 0^+_2$ transition in $^4$He exhibits good convergence at $\hbar\Omega=7$~MeV. Our results exceed the experimental data by a sizable amount and are comparable with the results of the EIHH calculations using the chiral $NN$ (N$^3$LO) + 3$N$ (N$^2$LO) interactions. The continuum is taken into account with the LIT in the EIHH approach while we treat the $0^+_2$ state as a bound state. The Daejeon16 interaction used in this work originates from the same N$^3$LO interaction utilized in the EIHH calculations and the 3$N$ force effects are mimicked in the Daejeon16 interaction by its off-shell features. Therefore, the remarkable consistency between our NCSM results and the EIHH results suggests that the coupling to the continuum may not be essential for the transition monopole form factor of $^4$He. From a comparison with all the available theoretical results  for this observable, we see that there is a strong dependence on the inter-nucleon interactions, as already indicated in Ref. \cite{Bacca:2012xv}. We demonstrate that although several calculations claimed to agree with the MAMI measurement, the EIHH is not the only realistic one that does not. We indicate that the gap between the {\it ab initio} calculation and the experimental data is not resolved
and deserves more investigations in the future. In particular, benchmark studies comparing the NCSM results with other theoretical approaches using the same inter-nucleon interactions would be valuable for understanding the discrepancies between the theoretical predictions and the experimental data.

\textit{Acknowledgments.}
We acknowledge helpful discussions with Ulf-G.~Mei\ss{}ner, Shihang Shen, Michele Viviani, Evgeny Epelbaum, Yong-Hui Lin, Pieter Maris and Nicolas Michel.
AMS is thankful for the hospitality of colleagues from Pacific National University (Khabarovsk, Russia) where a part of this work was done. This work is supported in part by the Ministry of Science and Higher Education of the Russian Federation (Project No. FEME-2024-0005).
A portion of the computational resources are provided by the National Energy Research Scientific Computing Center (NERSC), a U.S. Department of Energy Office of Science User Facility located at Lawrence Berkeley National Laboratory, operated under Contract No. DE-AC02-05CH11231 using NERSC awards NP-ERCAP0020944, NP-ERCAP0023866 and NP-ERCAP0028672.
This work is partially supported
by new faculty startup funding by the Institute of Modern Physics, Chinese Academy of Sciences, by Key Research Program of Frontier Sciences, Chinese Academy
of Sciences, Grant No. ZDB-SLY-7020, by the Natural
Science Foundation of Gansu Province, China, Grant
Nos.~20JR10RA067 and 23JRRA675, by the Foundation for Key Talents of Gansu Province, by the Central Funds Guiding the Local Science and Technology Development of Gansu Province, Grant No. 22ZY1QA006, by Natural Science Foundation of Henan under Grant No.252300421486, by international partnership program of the Chinese Academy
of Sciences, Grant No. 016GJHZ2022103FN, by National Natural Science Foundation of China, Grant No. 12375143, by National Key R\&D Program of China,
Grant No. 2023YFA1606903 and by the Strategic Priority Research Program of the Chinese Academy of Sciences, Grant No. XDB34000000.
P.~Yin and B.~Zhou are supported by Shanghai Research Center for Theoretical Nuclear Physics, NSFC and Fudan University, Shanghai 200438, China and the National Natural Science Foundation of China under Grant No.12147101.
A.~M.~Shirokov is thankful to the Chinese Academy of Sciences President's International Fellowship Initiative Program (Grant No.~2023VMA0013) which supported his visits to Lanzhou where a part of this work was performed and acknowledges the hospitality of Chinese colleagues during these visits. S.B. is supported by the Deutsche
Forschungsgemeinschaft
through the
Cluster of Excellence ``Precision Physics, Fundamental
Interactions, and Structure of Matter" (PRISMA$^+$ EXC 2118/1, Project ID 390831469) and through the Collaborative Research Center ``Hadron and Nuclei as discovery tools" (Project ID 514321794).
H.~Li is supported by CSC scholarship.
This material is based upon work supported by the U.S.~Department of Energy, Office of Science, under Award Nos.~DE-SC0023495 and DE-SC0023692. A portion of the computational resources were also provided by Gansu Computing Center and Gansu Advanced Computing Center.

\end{document}